\documentclass[prd,a4paper,
showpacs,
twocolumn
preprint
]{revtex4}
\def\eq#1{{Eq.~(\ref{#1})}}

\def\Tr{\mbox{Tr}\,}

\def\di{\mbox{d}}
\def\Mod{\mbox{mod}}
\newcommand{\be}{\begin{equation}}
\newcommand{\ee}{\end{equation}}
\newcommand{\bea}{\begin{eqnarray}}
\newcommand{\eea}{\end{eqnarray}}

\renewcommand{\Box}{\partial_{\mu}\partial^{\mu}}
\begin{document}
%%%%%%%%%%%%%%%%%%%%%%%%%%%%%%%%%%%%%%%%%%%%%%%%%%%%%%%%%%%  FRONT PAGE
\title{Axion and neutrino physics from anomaly cancellation}
\date{August 3, 2001}
\author{M.~Fabbrichesi}
\author{M.~Piai}
\author{G.~Tasinato}
\affiliation{INFN, Sezione di Trieste and\\
Scuola Internazionale Superiore di Studi Avanzati\\
via Beirut 4, I-34014 Trieste, Italy.}
\begin{abstract}
It has been recently shown that the requirement of anomaly cancellation
in a (non-supersymmetric)
six-dimensional version of the standard model
fixes the field content to the known three generations. 
We discuss the phenomenological consequences of
the cancellation of the local anomalies:
the strong CP problem is solved  and the  fundamental scale 
of the theory is bounded by the physics of the axion.
Neutrinos acquire a
mass in the range suggested by atmospheric experiments.
\end{abstract}
%\keywords{}
%\preprint{SISSA ??/00/EP}
%%%%%%%%%%%%%%%%%%%%%%%%%%%%%%%%%%%%%%%%%%%%%%%%%%%%%%%%%%%%%%%%%%%
\maketitle
%%%%%%%%%%%%%%%%%%%%%%%%%%%%%%%%%%%%%%%%%%%%%%%%%%%%%%%%%%%%%%%%%%%

\section{Introduction} 

Anomalies and their cancellation are crucial to
our understanding of
quantum gauge field theory. They can be used as a well-motivated
selection rule in fixing the field content of a model. 
The authors of a recent paper~\cite{Dobrescu:2001ae} consider the
standard model  in 
six space-time dimensions and show that 
it is possible to predict  the number of 
matter families (generations) by requiring the cancellation of a
global anomaly~\cite{Witten:1982fp}. The  cancellation of the gauge 
anomalies in six dimensions is achieved by means of the Green-Schwarz 
mechanism~\cite{Green:1984sg}.
In this paper we discuss some phenomenological consequences of
the model introduced in~\cite{Dobrescu:2001ae}. 

Axions are necessarily present
in the four-dimensional theory after the cancellation of the 
gauge anomalies in six dimensions.
They provide a solution to the strong CP problem. The  experimental constraints on  axion 
couplings yield  bounds on  the free parameters of the model, namely,
the compactification radius  $R^{-1} > 10^{6}$ TeV. At the same time, 
neutrinos acquire a mass of the right order of magnitude by means of 
a see-saw mechanism with the right-handed neutrinos required
by the cancellation of the gravitational anomaly in the six-dimensional theory.

We only consider  the minimal version
of the model with 
 two extra dimensions in which standard model fields 
are allowed to propagate: the field content of the theory is then 
completely determined by the requirement of anomaly cancellation,
and no other field, symmetry or additional extra-dimension is 
introduced.  

\section{Field content: anomaly cancellation}

Let us write a (non-supersymmetric)  six-dimensional theory, 
based on the standard model gauge group $SU(3)_c \times SU(2)_L \times U(1)_Y$;
matter fields are assigned 
to chiral fermions (projected by $(1 \pm \Gamma_7)/2$)
 in the usual representations of the gauge group
$Q$, $L$, $U$, $D$ and $E$, as shown in Tab.~\ref{fields}.
A scalar doublet $h$ must be present for the usual Higgs mechanism to 
take place. 
As explained in~\cite{Dobrescu:2001ae}, cancellation of purely 
gravitational and irreducible gauge anomalies forces 
quark singlets to have opposite chirality with respect to doublets,
and to introduce a standard-model singlet $N$ in order to have the same 
number of fermions of both chiralities.
Our (conventional) choice is to assign positive chirality to doublets 
and negative to singlets.  An alternative choice would be 
to assign opposite chiralities to leptons and hadrons: in
what follows this would turn in a change of ${\cal{O}}(1)$ in the  couplings
introduced to cancel anomalies, with minor modifications in the 
phenomenological consequences of the model.
%%%%%%%%%%%%%%%%%%%%% table
%%%%%%%%%%%%%%%%%%%%%%%%%%%%%%%%%%%%%%%%%%%%%%%%%%%%%%%%%%%%%%%%%%%
\begin{table}[h]
\begin{center}
\caption{Fermionic field content for each family. The six-dimensional
chirality is the eigenvalue of $\Gamma_7$.}
\label{fields}
%\begin{ruledtabular}
\vspace{0.2cm}
\begin{tabular}{|c|c|c|c|c|}
\hline \hline
 & \ \ Chirality \ \ & $\;\;\;\;U(1)_Y\;\;\;$ & $\;\;\;\;SU(2)_L\;\;$ & $\;\;\;\;SU(3)_c\;\;$ \cr
\hline
$\;\;\;\;Q\;\;\;\;\;$ & $+$ & $1/6$ & 2 & 3 \cr
$\;\;\;\;U\;\;\;\;\;$ & $-$ & $2/3$ & 1 & $\bar{3}$ \cr
$\;\;\;\;D\;\;\;\;\;$ & $-$ & $-1/3$ & 1 & $\bar{3}$ \cr
$\;\;\;\;L\;\;\;\;\;$ & $+$ & $-1/2$ & 2 & 1 \cr
$\;\;\;\;E\;\;\;\;\;$ & $-$ & $-1$ & 1 & 1 \cr
$\;\;\;\;N\;\;\;\;\;$ & $-$ & $0$ & 1 & 1 \cr
\hline \hline
\end{tabular}
%\end{ruledtabular}
\end{center}
\end{table}
%%%%%%%%%%%%%%%%%%%%%%%%%%%%%%%%%%%%%%%%%%%%%%%%%%%%%%%%%%%%%%%%%%%%%%%%%%%%%%%

Cancellation of global anomalies is obtained with $n_g = 0\; \Mod\; 3$ 
copies of this matter content. 
In particular  $n_g=3$ is in agreement with the experiment~\cite{Adeva:1992rp}.
We are thus left with Abelian and non-Abelian reducible gauge anomalies,
that would spoil unitarity unless cancelled by the Green-Schwarz 
mechanism. Accordingly, to recover non-Abelian gauge symmetries
we must introduce two real antisymmetric tensors $B^{L}_{M N}$ and 
 $B^{c}_{M N}$
whose couplings are
tuned to exactly cancel the anomalous terms in the gauge transformations.
The $B^{L,c}_{M N}$ fields transform in a non-standard way under gauge transformation:
\bea
\begin{array}{c}
\delta_{L}B^L  \equiv   - \frac{1}{\Delta_{L}^{ \;2}} \omega^L_2 =
- \frac{1}{\Delta_{L}^{ \;2}} \Tr \{ \beta\, \di W \}\, ,
\\
\delta_{c}B^{c}  \equiv  - \frac{1}{ \Delta_{c}^{ \; 2}} 
\omega^{c}_2 =
- \frac{1}{\Delta_{c}^{ \; 2}} \Tr \{ \gamma\, \di G \}\, . \label{1}
\end{array}
\eea
In \eq{1}, $\beta$ and $\gamma$ are 
the local parameters of $SU(2)_L$ and $SU(3)_c$
gauge transformations, while $\Delta_{L}$  and $\Delta_{c}$ are
free mass parameters.
 
The six-dimensional anomaly free Lagrangian can be written as:
\be
{\cal{L}}^{6D} = {\cal{L}}^{6D}_{SM}+{\cal{L}}^{6D}_{GS} \, ,\label{2}
\ee
where
\bea
\label{sixdim}
{\cal{L}}^{6D}_{SM} & = & \sum \bar{\psi} i \gamma^{M} D_{M} \psi 
 -\frac{1}{4} F_{M N} F^{M N} \nonumber \\ 
    &&   -\frac{1}{2} \Tr L_{M N} L^{M N}    -\frac{1}{2} \Tr V_{M N} V^{M N}  
\nonumber \\ 
    &&  +(D_{M} h)^{\dagger} D^{M} h   \;\;  + V(\phi^{\dagger}\phi)
\label{3}\\
    &&  + \Big(Y_l\, R\, \bar{L}h E + Y_d\, R\, \bar{Q}h D 
\nonumber \\
    &&  + Y_u\, R\, \bar{Q}\tilde{h} U + Y_{\nu}\, R\, \bar{L}\tilde{h} N
       \; + \; \mbox{h.c.}\Big)
\nonumber \\
    &&  + \frac{1}{2} M_N \left( N^T C^{-1} N - \bar{N} C 
 \bar{N}^T \right)
\nonumber 
\eea
and
\bea
{\cal{L}}^{6D}_{GS}  & = &   \frac{g^{\prime \; 3} R^3}{16
\pi^3}\sigma F_{MN} F_{RS} F_{PQ} \epsilon^{MNRSPQ}  \nonumber  \\
    &&  + \frac{g^{4} R^4\Delta_L^2}{6 \pi^3} B^L_{MN} \Tr\{ L_{RS} L_{PQ}\} \epsilon^{MNRSPQ} \nonumber \\
    &&  + \frac{ g^{2}g^{\prime \; 2}R^4\Delta_L^2}{144 \pi^3}
       B^L _{MN}  F_{RS} F_{PQ} \epsilon^{MNRSPQ} 
\nonumber \\
    &&  - \frac{ g^{2}g^{\prime \; }R^3 }{72 \pi^3}
       \sigma F _{MN} \Tr \{L_{RS} L_{PQ}\} \epsilon^{MNRSPQ} 
\nonumber \\
    &&  +\frac{ g_s^{2}g^{\prime \; }R^3 }{48 \pi^3}
       \sigma F _{MN} \Tr \{V_{RS} V_{PQ}\} \epsilon^{MNRSPQ} 
 \label{4} \\
    &&  - \frac{g_s^{2}g^{\prime \; 2}R^4 \Delta_{c}^{ \; 2}}
        {96 \pi^3}    B^{c} _{MN}  F_{RS} F_{PQ} \epsilon^{MNRSPQ}  
\nonumber \\
    &&  + \frac{ g_s^{2}g^{2}R^4 \Delta_{L}^{ 2} }{48 \pi^3}
      B^{L} _{MN}  \Tr\{ V_{RS} V_{PQ}\} \epsilon^{MNRSPQ}  
\nonumber \\
    &&  + \frac{g_s^{2}g^{2}R^4 \Delta_{c}^{ \; 2}}
        {48 \pi^3}    B^{c} _{MN} \Tr \{ L_{RS} L_{PQ}\} \epsilon^{MNRSPQ}  
\nonumber \\
    && +\frac{1}{12} H^{L,c}_{M N S} H^{L,c \;M N S}\;.
\nonumber
\eea
In \eq{3}, $\psi$ is a generic fermionic chiral fields,  $D_M$ the covariant 
derivative on the associated gauge representation, 
$V_{M N}$, $L_{M N}$ and $F_{M N}$ are the field 
strength tensors of the gauge bosons $G_M$, $W_M$ and $A_M$
of $SU(3)_c$, $SU(2)_L$ and $U(1)_Y$ respectively.
${\cal{L}}^{6D}_{SM}$ is the usual standard model Lagrangian,
with Lorentz indexes in six dimensions, to which we are 
allowed to add a Yukawa 
interaction also for neutrinos ($\tilde{h} = i\sigma_2 h^{\ast}$) 
and a Majorana
mass term for the singlet $N$, which behaves like a right-handed neutrino.
The scalar potential $V$ is a power series in
$\phi^{\dagger}\phi$. The coefficients of the Green-Schwarz terms in
\eq{4} match the one-loop anomalous terms,  computed for six
space-time dimensions in~\cite{Frampton:1983ah}.

The Lagrangian in \eq{4} contains  the (gauge non-invariant) terms required 
for the cancellation of all 
reducible gauge anomalies, and the kinetic terms for 
the 2-forms $B^{L,c}_{MN}$, with:
\be
H^{L,c} \equiv \di B^{L,c} - 
\frac{1}{\Delta_{L,c}^{\;2}} \omega_3^{L,c}\, ,
\ee
and
\bea
\begin{array}{ccc}
\omega_3^{c}  & =  & \Tr \{G \wedge V -\frac{1}{3}g_s R\; 
   G \wedge G \wedge G \}\;, 
\\
\omega_3^{L} & = & \Tr \{W \wedge L -\frac{1}{3}g R\; W \wedge W \wedge W \}\, . \end{array}
\eea
The Chern-Simons forms $\omega_3^{c,L}$ are needed
to make $H_{MNS}^{L,c}$ invariant, and  satisfy the relations
$\delta_{L,c}\omega_3^{L,c} = -\di \omega_2^{L,c}$~\cite{Zumino:Ed.ew}.  

The presence of the scalar field $h$ can be used 
to cancel the $U(1)$ anomalies.
The Higgs field has been decomposed in a doublet $\phi$ 
with vanishing hypercharge and a $SU(2)_L$ singlet $\sigma$ writing:
\bea
h = \phi\, e^{i \sigma}\;.
\eea
Under a $U(1)_Y$ gauge transformation 
$\delta \sigma = \frac{1}{2} g^{\prime} R \alpha$, being $\alpha$ the 
parameter of the transformation. In unitary gauge $\sigma=0$, 
the gauge bosons acquire mass in the standard way and terms proportional to $\sigma$ in \eq{4} vanish. 

We have not written explicitly the terms that are needed to cancel 
mixed (gauge-gravitational) anomalies, because they are not 
relevant for the phenomenological discussion we are interested in.
To achieve the cancellation it is enough to add for each gauge group
couplings of the form
\bea
\omega_3\, \Omega_3 + B\,  \Tr  {\cal{R}}\wedge {\cal{R}} \, ,
\eea
with appropriate coefficients, where $B$ are 
the antisymmetric tensors $B^{L}$, $B^{c}$ and $\sigma \, F$,
$\Omega_3$ is the gravitational Chern-Simon form defined by 
$\di \Omega_3= \Tr {\cal{R}}\wedge {\cal{R}} $. ${\cal{R}}$ is the Ricci tensor.
For details see~\cite{Frampton:1985wc}.

We denote by $R^2$ the volume of the compact extra-dimensions, so that
the Newton constant is related to the fundamental scale $M_f$
of the theory by 
\bea
M_{Pl} = R M_f^2\;.  \label{planck}
\eea
By  writing  the
dimensionfull couplings as $gR$ yields, after dimensional reduction,  
the (four dimensional) gauge couplings $g_s$, $g$ and 
$g^{\prime}$.
The only remaining parameters in the Lagrangian are 
the mass parameters $M_N$, $\Delta^{L}$, $\Delta^{c}$ and
the couplings of the scalar potential.

\section{Dimensional reduction}

The two extra dimensions are assumed to be compact, and the underlying 
geometry flat. 
Chiral fermions in six dimensions correspond to Dirac fermions in
four dimensions, but chirality is recovered by orbifold projection. 
We assume space-time to be
\bea
{\cal{M}}_4 \times \frac{S^1 \times S^1}{Z_2}\, , 
\eea
the product of four-dimensional Minkowski and a torus with orbifold
$Z_2$ which impose a symmetry  under the parity transformation
\be
Z_2 :  (y,z) \rightarrow (-y,-z)\, , 
\ee
where $(y,z)$ are the coordinates on
the torus $S^1 \times S^1$.

In what follows, we assume that we are allowed to work in the 
limit of dimensional reduction,
in which the low-energy Lagrangian contains only the zero 
modes of the fields---while higher modes decouple because of 
their large masses, 
proportional to $2 \pi /R$.
The effects of this simplifying assumption should be checked at the end
for consistency to make sure that the large number of these heavier
states do not enhance potentially dangerous operators. 

A consistent assignment of
$Z_2$-parities makes it possible  to have a single massless
chiral field out of each 
$\psi$; the projection gives a factor $1/2$ in  the Green-Schwarz 
gauge non-invariant terms in \eq{4}.  The reduced Lagrangian also 
contains the zero modes of $h$ and of the gauge bosons, together
with two anti-symmetric tensors $B^{L,c}_{\mu \nu}$, and two pseudo-scalars
$b^{L,c}$, coming, respectively, from the $\{0123\}$ and $\{56\}$ sectors,
of the decomposed tensors 
\be
B^{L,c}_{MN} \rightarrow 
b^{L,c}\equiv \sqrt{3}/6\,\epsilon^{\hat{M}\hat{N}}\,
B^{L,c}_{\hat{M}\hat{N}} \quad  \hat{M}\hat{N} = 5,6 \, .
\ee 
There is no zero mode for the $\{56\}$ part of 
the field strength tensors.
In four dimensions an antisymmetric tensor is equivalent to a 
pseudo-scalar, and we redefine:
\bea
\partial_{\mu}c^{L,c} & \equiv & i\frac{1}{6}
\epsilon_{\mu}^{\;\;\nu\rho\sigma}H^{L,c}_{\nu\rho\sigma} \, .
\eea
We use Greek indexes for 4-dimensional quantities.

The spectrum, after integrating out the compact dimensions, is 
the same as in standard model, with four additional pseudo-scalar fields   
$c^L$, $c^{c}$, $b^L$ and  $b^{c}$~\cite{Witten}, and the Lagrangian becomes:
\bea
{\cal{L}}^{4D}& =& {\cal{L}}^{SM}
        +  (Y_{\nu} \bar{L}\tilde{H} N + \mbox{h.c.})
\nonumber \\
    &&  + \frac{1}{2} M_N (N^T C^{-1} N - \bar{N} C \bar{N}^T)
\nonumber \\
    &&  + \frac{g^{4}R^3\Delta_{L}^{\;2}}{6 \sqrt{3} \pi^3} b^L \; \Tr L\tilde{L} 
     + \frac{g_s^{2}g^{2}R^3 \Delta_{c}^{ \; 2}}
        {48 \sqrt{3} \pi^3}    b^{c} \; \Tr  L\tilde{ L} 
\nonumber \\
&&    + \frac{ g^{2}g^{\prime \; 2}R^3\Delta_{L}^{\;2}}{144\sqrt{3} \pi^3}
       b^L \; F\tilde{ F} 
      - \frac{g_s^{2}g^{\prime \; 2}R^3 \Delta_{c}^{ \; 2}}
        {96 \sqrt{3} \pi^3}    b^{c}\;   F\tilde{F} 
\nonumber   \\
&&  + \frac{ g_s^{2}g^{2}R^3 \Delta_{L}^{\; 2} }{48 \sqrt{3} \pi^3}
      b^L \; \Tr V\tilde{ V }
\label{4d}\\
   && + \frac{1}{2} \partial_{\mu}b^L\;\partial^{\mu}b^L +
\frac{1}{2} \partial_{\mu}b^{c}\;\partial^{\mu}b^{c}
\nonumber \\
  && + \frac{1}{2} \partial_{\mu}c^L\;\partial^{\mu}c^L +
\frac{1}{2} \partial_{\mu}c^{c}\;\partial^{\mu}c^{c} \nonumber
 \\
  && - \frac{c^L}{3 \Delta_{L}^{\;2} R} \Tr L \tilde{ L}
 - \frac{c^{c}}{3 \Delta_{c}^{\;2} R} \Tr V \tilde{ V} \, .
\nonumber 
\eea
The Lagrangian ${\cal{L}}^{SM}$ in \eq{4d}
 is  the standard model Lagrangian (
in the unitary gauge).
The last terms have been obtained using the (six dimensional) 
 Bianchi identity
\be
\label{defH}
\di H^{L,c} = -\frac{1}{\Delta_{L,c}^{ \;2}} \di \omega^{L,c}_3\; ,
\ee
that in $D=4$ gives the equation of motion for the fields 
$c^L$ and $c^c$:
\bea
\Box c^L &=& - \frac{1}{3 \,\Delta_{L}^{ \;2} R} \Tr L \tilde{L}\, ,
\label{mot}\\
\Box c^{c} &=& - \frac{1}{3 \,\Delta_{c}^{\;2} R} \Tr V \tilde{V}\, .
\nonumber
\eea

Only three linear combinations $\varphi_1$, $\varphi_2$ and
$\varphi_3$ of the  
four pseudo-scalars are coupled to 
 gauge fields by means of axion-like terms, 
while the orthogonal combination 
gives rise to a massless free field with no phenomenological 
consequence. 

\section{Axions}

After removing the  decoupled scalar from  the Lagrangian in \eq{4d}, we obtain
\bea
{\cal{L}}^{4D}& =& {\cal{L}}^{SM}
        +  (Y_{\nu} \bar{L}\tilde{H} N + \mbox{h.c.}) \label{reduced}
\\
    &&  + \frac{1}{2} M_N (N^T C^{-1} N - \bar{N} C \bar{N}^T)
\nonumber \\
  && + \frac{1}{2} \partial_{\mu}\varphi_1\;\partial^{\mu}\varphi_1 +
\frac{1}{2} \partial_{\mu}\varphi_2\;\partial^{\mu}\varphi_2
 +
\frac{1}{2} \partial_{\mu}\varphi_3\;\partial^{\mu}\varphi_3
\nonumber \\
    &&  + \varphi_1 \; \left[ \frac{1}{F_1^F} \,F\tilde{F} + 
\frac{1}{F_1^L} \, \Tr\,L\tilde{L} + \frac{1}{F_1^V} \, \Tr\,V\tilde{V} \right] \nonumber \\
      & &            + \varphi_2   \; \left[ \frac{1}{F_2^F} \,F\tilde{F} + 
 \frac{1}{F_2^V} \, \Tr\,V\tilde{V} \right] +
\varphi_3   \;  \frac{1}{F_3^V} \, \Tr\,V\tilde{V}\, ,
\nonumber
\eea
where the constant $F_i^{F,V,L}$ are functions of the coefficients in front of the scalar-gauge fields coupling terms in \eq{4d}.
The fields $\varphi_i$ have the same couplings 
of the Peccei-Quinn  axion: 
they are invariant under translations but for the coupling to the 
gauge fields.  

The axion 
solves the strong CP problem~\cite{Peccei:1977hh,Wilczek:1978pj,Kim:1979if}:
a term  in the form
\bea
\bar{\theta}\, \frac{\alpha_s}{8\pi}\,\Tr V \tilde{V}\, ,\label{T}
\eea
which is allowed by the symmetries of non-Abelian gauge theories,
would 
induce an electric dipole moment  for neutrons in conflict with 
experimental data
unless $\bar{\theta} < 10^{-14}$. 

It is not possible to put by hand $\bar{\theta} = 0$, because of
instantonic effects, 
but the addition of a pseudo-scalar 
field $a$ with coupling
\bea
\frac{\alpha_s}{8\pi}\,\frac{1}{F_a}\,a\, \Tr V \tilde{V}\;
\eea
and with no tree-level potential gives dynamically 
\bea
\frac{\langle a \rangle}{F_a}+\bar{\theta}=0\, .
\eea
In this way, no electric dipole moment is generated, and 
CP is restored as a good symmetry of the QCD Lagrangian.

There are experimental constraints on the axion couplings coming from 
the combination of cosmological, astrophysical and accelerator 
searches~\cite{PDG}. 
In order to perform the comparison with the experimental constraints, it is 
necessary to write down the low-energy effective theory in terms of
photons, pions, nucleons and axions only. 

In the low-energy theory, we can safely neglect interactions 
with $Z$ and $W$ bosons, and extract only the  
electromagnetic couplings
\be
\left\{ \begin{array}{ccc}
           \Tr L \tilde{L} & = & \frac{1}{2}\, \sin^2\theta_{W} \;F_{em} \tilde F_{em} + \cdots \\
            F \tilde{F} & = &  \cos^2\theta_{W}\; F_{em} \tilde F_{em} + \cdots \, ,
\end{array} \right. \label{em}
\ee
after the rotation of neutral bosons by the weak angle $\theta_W$.
Accordingly, only two combinations out
 of the three $\varphi_i$ fields couple to the massless gauge fields: 
one to photons and gluons, the other to photons only.
The former has the correct couplings and transformation 
properties to be identified with the Peccei-Quinn axion. Its presence
is a consequence of the anomaly cancellation, and therefore
of the choice of writing a six-dimensional gauge theory.

Now we turn the coupling  to gluons into a
coupling to quarks. This can be achieved by a chiral transformation.
Then, using the methods of current algebra, we rewrite
the theory in terms of pions, and eliminate quarks and gluons.

Adding the coupling of pions to photons, responsible
for the decay $\pi^0 \rightarrow 2\gamma$, yields the
interaction terms needed to compute all the contributions
to the mass matrix of pions and axions.
After all of these manipulations we can write
\bea
{\cal{L}}^{\pi} & = & \frac{1}{2}\partial_{\mu}\pi^0 \partial^{\mu}\pi^0
      +  \frac{1}{2}\partial_{\mu}a\, \partial^{\mu}a
    +  \frac{1}{2}\partial_{\mu}\, a^{\prime}
 \partial^{\mu}a^{\prime}
\nonumber \\
  & - &   \frac{1}{2} \left(\begin{array}{ccc} \pi^0 & a & a^{\prime}  \end{array} \right) \, {\cal{M}}^2 
\left(\begin{array}{c}\pi^0 \cr a  \cr a^{\prime}  \end{array} \right)
 \\
  & + &  \Big(\frac{\pi^0}{f_{\pi}}+\frac{a}{f_a}+\frac{a^{\prime}}
{f_{a^{\prime}}}
      \Big)\,\frac{\alpha}{8 \, \pi}
	F_{e.m.}^{\mu \nu} F_{e.m.}^{\rho \sigma} \, 
        \epsilon_{\mu \nu \rho \sigma} \nonumber
\eea
where
\begin{widetext}
\be
 {\cal{M}}^2   = \quad\quad\Delta m^2_{\pi} \; \left(\begin{array}{ccc}
   m^2_{\pi^0}/ \Delta m^2_{\pi} \quad\quad & \quad\quad  f_{\pi}/f_a \quad\quad & \quad   k + f_{\pi}/f_{a^\prime} 
\\
 f_{\pi}/f_a &   \left( f_{\pi}/f_a \right)^2 & 
  f_{\pi}^2/(f_a f_{a^\prime})
\\
 k +  f_{\pi}/f_{a^\prime} & f_{\pi}^2/(f_a f_{a^\prime}) &  m^2_{\pi^+}/ \Delta m^2_{\pi} ( f_{\pi}/2m)^2 +  \left( f_{\pi}/f_{a^\prime} \right)^2
\end{array} \right) \, .
\ee
\end{widetext}
The parameter $m\equiv F_3^V (\alpha_s/ 4 \pi)$ and
\be
k \equiv \frac{m_{\pi^+}^2}{ \Delta m^2_{\pi}} \frac{m_d-m_u}{m_d+m_u} \frac{f_\pi}{2 m} \, .
\ee
The masses $m_u$ and $m_d$ are those of up and down quarks,
$f_{\pi} \simeq 93$ MeV is the pion decay constant, 
$m_{\pi^0}$  and $m_{\pi^+} $  are the masses
of the pions, while $\Delta m_{\pi}^2 \equiv
m_{\pi^0}^2-m_{\pi^+}^2$. 
The decay constants are normalized in such a way that the 
partial decay rate of neutral pions into photons is
\be
\Gamma(\pi^0\rightarrow 2\gamma) =  \frac{\alpha^2 
\,m_{\pi}^3}{64\pi^3\,f_{\pi}^2}
\simeq 7.6\; \mbox{eV}\, .
\ee
The partial diagonalization of this matrix makes it possible to 
identify the physical
pion field and the couplings of the two remaining light pseudoscalars
$a$ and $a^\prime$.
The coupling of the axions to the photon comes both 
from \eq{em} and pion-axion mixing.
A stringent experimental bound to consider comes from helium 
burning lifetimes of red giants, and  imposes an upper limit to
the coupling axion-photon~\cite{Raffelt:1986nk}
\be 
  g_{a\gamma} < 10^{-10} \;\mbox{GeV}^{-1} \label{bound}
\ee
with
\be
{\cal L} = - \frac{1}{4} g_{a\gamma} \,a\; F^{\mu\nu}\tilde F_{\mu\nu} \, .
\ee
The limit of vanishing
masses for axions can be used.  
In our case, we have that
\begin{widetext}
\be
g_{a\gamma} \equiv \frac{\alpha}{\pi} \; 
\sqrt{
\left[ 
\frac{1}{f_a}  
\left( 
1+ \frac{\Delta m^2_{\pi}}{m^2_{\pi^0}} 
\right) 
\right]^2
+\left[
\frac{1}{f_{a^{\prime}}}
\left( 
1 + \frac{\Delta m^2_{\pi}}{m^2_{\pi^0}} + k \frac{f_{a^\prime}}{f_\pi} \frac{\Delta m^2_{\pi}}{m^2_{\pi^0}}  
\right) 
 \right]^2 
 } 
\, .
\label{26}
\ee
\end{widetext}
The coupling $g_{a\gamma}$ depends both on positive and negative 
powers of  $\Delta_c$ and $\Delta_L$---through the parameters $m$,
$f_a$  and $f_{a^\prime}$ in \eq{26}, which, in turns, come from the
couplings in \eq{4d}. For a fixed value of the radius
$R$, there exists a minimum of  $g_{a\gamma}$ as a function of
these free  parameters. Taking this minimum and comparing it with the 
bound in \eq{bound}, yields a constraint on the possible values of $R$. 
 
A similar bound is obtained by considering the coupling of axions to nucleons
\be
{\cal L } = -i g_{aN} \,\bar N \gamma_5 N \, a \, ,
\ee
where, in our case
\begin{widetext}
\be
g_{aN} = \frac{g_A m_N}{2} 
\sqrt{
\left[ 
\frac{1}{f_a}  
 \frac{\Delta m^2_{\pi}}{m^2_{\pi^0}}  
\right]^2
+\left[
\frac{1}{f_{a^{\prime}}}
\left( 
 \frac{\Delta m^2_{\pi}}{m^2_{\pi^0}} + k \;\frac{f_{a^\prime}}{f_\pi} \frac{\Delta m^2_{\pi}}{m^2_{\pi^0}}  
\right) 
 \right]^2 
}
 \, , \label{su}
\ee
\end{widetext}
and $g_A$ is the axial nucleon coupling, whereas $m_N$ is the nucleon
 mass. Equation (\ref{su}) is obtained by including only the mixing 
between the neutral pion and the axion.

Limits coming from supernova SN1987a~\cite{Raffelt:1990yz} impose
\be
 g_{aN} < 3 \times 10^{-10} \, . \label{N}
\ee

The two bounds \eq{bound} and \eq{N} give
\be
\frac{1}{R}  >   10^6 \; \mbox{TeV}\,\label{R} ,
\ee 
which, applying \eq{planck}, corresponds to
\be
M_f >   10^{11}\; \mbox{TeV}\, . \label{M}
\ee
We have thus obtained an explicit lower limit on the fundamental scale
from the experimental bounds on axion couplings.

\section{Neutrinos} 

Let us recall that right-handed neutrinos must be included in 
six dimensions in order to cancel the gravitational anomalies. 
The Lagrangian in \eq{reduced} has two mass terms for the neutrinos:
\be
 \frac{1}{2} M_N \left( N^T C^{-1} N - \bar{N} C 
 \bar{N}^T \right) 
\ee
and the Dirac mass term induced by the Yukawa coupling
after electroweak symmetry breaking. The latter has the same structure 
of the other fermion masses $m_{\nu}^D \equiv \langle \phi^0\rangle \,Y_{\nu}
= v Y_{\nu}/\sqrt{2}$.
Together they give rise to a neutrino mass matrix 
\be
\left( \begin{array}{cc} 0 & m_\nu^D \\
                                    m_\nu^D & M_N
        \end{array} \right) \, ,
\ee
which is of the right form for the see-saw mechanism~\cite{Gell-Mann:1980vs}.
Since there is no symmetry 
to protect $M_N$, the right-handed Majorana mass term,
it is reasonable to assume that $M_N \sim M_f$.  Accordingly
  the mass of the light neutrinos is given  by the see-saw expression:
\bea
m_{\nu} \sim \frac{(m_{\nu}^D)^2}{M_f} \, .
\eea
Imposing the heaviest mass to be the one measured by 
atmospheric neutrino experiments, $\sqrt{\Delta m_{atm}^2} \sim (0.04 \div 0.09)$ eV~\cite{Sobel}, we can estimate the 
required value for the Dirac mass term $m_{\nu}^D$ 
for the lightest allowed $M_f$:
\be
m_{\nu}^D \sim (65 \div 100)\; \mbox{GeV} \, ,
\ee
which is consistent with the usual mass terms for fermions.

\section{Higher-order operators} 

The model under consideration 
is non-renormalizable; it must be understood as the 
low-energy limit of a more fundamental theory which gives 
additional interactions above the cut-off scale  $M_f$.
These interactions  give rise to operators suppressed by 
powers of $1/M_f$ that violate
the global symmetries of the low-energy theory. However, because
of the  limit we obtained for $M_f$, these effects are
less worrisome than in models with large extra-dimensions
in which the typical scale of such operators is in the TeV range. 
Nevertheless, some potentially dangerous operators must be checked.
In particular,  operators like
\be
{\cal L} \sim \frac{1}{M_P^2}  Q Q  Q L \, ,
\ee
could lead to too fast a proton decay unless  $M_P$ is taken of 
the order of $10^{16}\,$GeV. They are,
however, excluded by the residual discrete symmetries 
that remain after compactification from the $SO(5,1)$ Lorentz 
symmetry in six dimensions~\cite{Appelquist:2001jz}. 

Operators compatible with these discrete symmetries could, for an 
arbitrary phase in the coupling, lead to potentially 
dangerous electric dipole moments
\be
{\cal L} = i\,e\, \frac{m_{\psi}}{M_d^2} 
\bar \psi \sigma^{\mu\nu} \psi F_{\mu\nu} \, .
\ee

Comparing $d \equiv e\, m_{\psi} /M_d^2$ with 
the experimental bound~\cite{ddee}
\bea
\left.
\begin{array}{c}
d_e < 2\times 10^{-27}\, \mbox{e\,cm}\; ,
\end{array}
\right.
\eea
we find 
\bea
\label{dipole} 
M_d^2 > 10^{4} \, \mbox{TeV}^2\; ,
\eea
which is satisfied by several orders of magnitude for $M_d \sim M_f$ imposing 
the bounds of \eq{R} and \eq{M}.
The similar flavor violating operator
\be
{\cal L} = i\,e\, \frac{m_{\mu}}{M_{\mu}^2} 
\bar e \sigma^{\mu\nu} \mu F_{\mu\nu} \, ,
\ee
would induce the decay $\mu \rightarrow e \gamma$ with
the partial rate
\be
\Gamma(\mu \rightarrow e \gamma) = \frac{\alpha 
m^5_{\mu}}{M_{\mu}^4}\; ,
\ee
where $m_{\mu}$ is the muon mass.
Comparing this with the experimental constraint~\cite{PDG},
\be
\Gamma(\mu \rightarrow e \gamma) < 4 \times 10^{-33}\, \mbox{TeV}\;,
\ee 
yields
\bea
\label{muon} 
M_{\mu}^2 > 10^{5} \, \mbox{TeV}^2\; .
\eea

Another class of potentially dangerous corrections comes from 
Kaluza-Klein states.  The bounds
we obtain for the extra-dimensional volume 
justifies the approach of working with only 
the zero modes of the theory: the first Kaluza-Klein excitations
are at a scale much larger than that experimentally relevant and
they can be safely neglected in the computation of observable quantities.  
Those processes that take place in the standard model only at the
one-loop level could be an exception to this conclusion. However, no
relevant effect is expected
for our value 
of the compactification radius (see, for instance,~\cite{Delgado:2000sv}).

\section{Discussion} 

We have discussed the phenomenological consequences 
of a six-dimensional realization
of the standard model. 

The cancellation of a global anomaly 
imposes the presence of three generations~\cite{Dobrescu:2001ae}. 
Local anomaly cancellation requires that  the
four-dimensional spectrum contains, besides
the usual fields of the standard model, 
right-handed neutrinos and axion fields. 
The axion fields solve the strong CP problem.

The fundamental scale of the theory is related to the decay constants
of the axions;  therefore, its value   must  be large enough to evade 
experimental bounds. The fundamental scale is thus bounded.
A problem of naturalness remains because of the large scale $M_f$ of
the theory: the Higgs sector requires fine-tuning in order to protect the 
weak scale. A dynamical explanation of the large difference between 
the electroweak symmetry breaking
scale and the fundamental scale is required because a supersymmetric version 
of the model has been shown to contain irreducible 
anomalies~\cite{Dobrescu:2001ae}.

Without any further assumptions, a  see-saw mechanism induced by the large mass
scale of the theory provides in a natural manner
a neutrino mass in the range indicated by atmospheric experiments.

In order to evade the bounds on axion physics, it has been suggested
to add mass terms localized at the fixed points of the orbifold
for the pseudoscalars~\cite{Dobrescu:2001ae}. 
Massive pseudoscalars are not axions: they
cannot be used to solve the strong CP problem, because of the lack of 
translational invariance. If heavy enough, they decouple from 
the low-energy phenomenology, and limits from axion physics do not
apply.
Such a term for 
$b$ and $b^{\prime}$ is forbidden by gauge invariance of the 
six-dimensional theory.
If it is possible to write it for  $c$ and $c^{\prime}$, 
by changing \eq{defH} and \eq{mot},
still the strong CP problem is solved, thanks to the fields
$b$ and $b^{\prime}$. Their
couplings involving only positive powers of $\Delta_L$ 
and $\Delta_c$, no bound can be deduced for
$R$. In particular $1/R \sim $TeV can be compatible with experiments, 
choosing  $\Delta_L \sim  \Delta_c \sim 10 $ GeV.

Nevertheless, while a softening of the hierarchy problem in
the Higgs sector would be accomplished in this way, translating it into
the dynamical problem of understanding the big difference between the scale 
of the large extra-dimensions and the fundamental scale of the theory,
the lowering of this scale 
would require unnaturally small couplings in higher order operators
and  loop dominated processes, giving potentially large contributions
to electroweak precision observables, flavor changing 
processes~\cite{Delgado:2000sv}  and CP 
violating quantities as electric dipole moments (see \eq{dipole}).

Furthermore, the good prediction
for neutrino masses would be lost, and their smallness in 
comparison to the other fermions would
again be unnatural, in presence of TeV scale 
right-handed neutrinos.

\acknowledgments

We thank L.~Bonora and M.~Serone  for useful discussions. This work is
 partially supported by  the European TMR Networks HPRN-CT-2000-00148 
and HPRN-CT-2000-00152.

\vspace{0.5cm}
%%%%%%%%%%%%%%%
%------------------------- REFERENCES ------------------------------

%\begin{references}

%\end{references}
\end{document}